\documentstyle[psfig]{article}
\begin{document}
\title{Beta-function in QCD  and  
gluon condensate}
\author{A.I. Alekseev\footnote{Electronic address: 
alekseev@mx.ihep.su}\\
{\it Institute for High Energy Physics, 142281 Protvino, 
}\\ {\it Moscow Region, Russia}\\
B.A. Arbuzov\footnote{Electronic address: 
arbuzov@theory.sinp.msu.ru} \\
{\it Skobeltsyn Institute for Nuclear Physics, Moscow 
State University,}\\ {\it 119899 Moscow, Russia}}
\date{}
\maketitle
\begin{abstract}
Under assumption of singular behavior of 
$\alpha_s(q^2)$ at $q^2 \simeq 0$ and of large $q^2$ 
behavior, corresponding to the perturbation theory 
up to four loops, a procedure is considered of 
matching the $\beta$-function at a boundary of 
perturbative and non-perturbative regions. The 
contribution of the non-perturbative region to 
the gluon condensate is calculated with varying  
normalization condition $\alpha_s(m^2_{\tau}) = 
0.29,\,0.30,\,...,\,
0.36$ for two different ways of definition of the 
non-perturbative invariant charge  
in the infrared region. We obtain quite consistent 
results for values of the gluon condensate, nonperturbative
region scale $q_0$, and the string tension $\sigma$. 
\end{abstract}

\bigskip

It is well-known, that the $\beta$-function in the 
perturbative QCD is of the form
\begin{equation}
\beta_{\rm pert}(h)=-b_0 h^2-2b_1h^3-
\frac{b_2}{2}h^4-b_3h^5
+O(h^6), \,\,\, h=\frac{\alpha_s(q^2)}{4\pi}.
\label{1}
\end{equation}
For $n_f=$ 3 we have values of coefficients $b_0=$ 9,  
$b_1=$ 32,  $b_2\simeq$
1287.67,  $b_3\simeq$ 12090.38 
(coefficients $b_0$,  
$b_1$ do not depend on renormalization scheme while 
values $b_2$, $b_3$  correspond to a choice of 
$\overline{MS}$-scheme). Expressions obtained by 
solution of Gell-Mann -- Low equation
\begin{equation}
q^2\frac{\partial h(q^2)}{\partial q^2}=\beta(h)
\label{2}
\end{equation}
with the use of~(\ref{1}), are widely used for 
sufficiently large momenta transfer, however they 
can not be applied in the infrared region. For 
a behavior of the invariant charge (the running coupling 
constant) $\alpha_s(q^2)$
at $q^2\rightarrow 0$ a number of variants are 
considered (see, e.g.,~\cite{ShirTMF02}). In particular singular 
infrared asymptotic at $q^2=0$ 
\begin{equation}
\alpha_s(q^2)\simeq \frac{g^2}{4\pi}\frac{M^2}{q^2}, \;\;\;
q^2\rightarrow 0
\label{3}
\end{equation}
is possible (see, e.g., review~\cite{Arb1} and more recent 
papers~\cite{AA}). 
Such behavior naturally corresponds to a linear confining 
quark-antiquark static potential at that $g^2 M^2\,=\,6\pi\sigma$,
where $\sigma$ is the string tension.
Results of some works on the lattice study 
of the three-gluon vertex~\cite{lat}  demonstrate 
a necessity of taking into account of non-perturbative 
contributions to the running coupling constant being 
of the form of~(\ref{3}). 
In the framework of the continuous QFT
arguments in favor of behavior~(\ref{3}) are also presented in
recent paper~\cite{Gogohia}. 

Asymptotic 
behavior~(\ref{3}) occurs provided 
$\beta(h)\rightarrow-h$ for $h\rightarrow \infty$. 
Let us consider a possibility of behavior~(\ref{3}) to 
be valid and assume the following form of the infrared 
$\beta$-function
\begin{equation}
\beta(h)=-h+z, \;\;\; h>h_0,
\label{4}
\end{equation}
where $z$ is a constant and $h_0$ defines the boundary 
between  perturbative and  non-perturbative regions. 
For $h<h_0$ we shall use $\beta$-function~(\ref{1}) with 
a finite number of loops taken into account. Our recipe for
construction 
of $\beta$-function for all $h>0$ consists in 
a smooth matching of expressions~(\ref{4}) and (\ref{1}) 
at point $h=h_0$ in approximations of the perturbation theory up to 
four loops. The demand of the $\beta$-function 
and its derivative to be continuous uniquely fix 
free parameters $z$ and $h_0$ of the ``global" 
$\beta$-function (the matched one). 

For an illustration let us consider the most simple 
one-loop case. Conditions of matching give two equations
\begin{eqnarray}
-b_0h_0^2&=&-h_0+z,
\nonumber \\
-2b_0h_0&=&-1.
\label{5}
\end{eqnarray}
The solution of set~(\ref{5}) reads 
\begin{equation}
h_0=\frac{1}{2b_0}, \;\;\; z=\frac{1}{4b_0}.
\label{6}
\end{equation}
We shall normalize perturbative solution 
\begin{equation}
\alpha_s(q^2)=\frac{4\pi}{b_0\ln x}, \;\;\;
x=\frac{q^2}{\Lambda^2_{QCD}}, \;\;\; q^2\ge q_0^2
\label{7}
\end{equation}
by value $4\pi h_0$, that gives
\begin{equation}
x_0=q^2_0/\Lambda^2_{QCD}=e^2,
\label{8}
\end{equation}
where $e = 2.71828...$ . Imposing on $\alpha_s(q^2)$ the 
natural condition to be continuous at $q^2=q^2_0$, we may 
normalize non-perturbative solution of equation~(\ref{2})
\begin{equation}
\alpha_s(q^2)=4\pi\left(\frac{C}{q^2}+z\right), \;\;\;
q^2\le q_0^2
\label{9}
\end{equation}
by $4\pi h_0$ as well. As a result we obtain
\begin{equation}
C=q_0^2(h_0-z), \;\;\; c_0\equiv C/\Lambda_{QCD}^2=x_0(h_0-z)=
\frac{e^2}{4b_0}.
\label{10}
\end{equation}
For final fixation of the solution for all $q^2>0$ we 
need to 
define $\Lambda_{QCD}$ by normalizing the solution, say, 
at point $q^2=m_{\tau}^2$, where $m_{\tau}=$ 1.77703 GeV
is the mass of the $\tau$-lepton~\cite{Data}.

Let us turn to  calculation of the gluon condensate. 
Its value is defined by the non-perturbative part of   
$\alpha_s$. We have (see, e.g., the third of 
refs.~\cite{AA}) 
\begin{equation}
K\equiv <\alpha_s/\pi: G^a_{\mu\nu} G^a_{\mu\nu}:>=
\frac{3}{\pi^3}\int\limits_{0}^{\infty} dq^2\,q^2 \alpha_{\rm
npt}(q^2)=
\frac{3}{\pi^3}\int\limits_{0}^{\infty} dq^2\,q^2(\alpha_s(q^2)-
\alpha_{\rm pert}(q^2)).
\label{11}
\end{equation}
In our approach the non-perturbative contribution is 
present for $q^2<q_0^2$ only. For the beginning we define 
the perturbative 
part at this region basing on the assumption of 
freezing of  $\alpha_s$ at small $q^2$ (see, 
e.g.,~\cite{Simonov}). That is we assume
 \begin{equation}       
\alpha_{\rm pert}(q^2)\,=\, 
\alpha_s(q_0^2)\,=4\pi h_0, \,\,\, q^2<q_0^2.
\label{12}
\end{equation}
Using expressions~(\ref{9}), (\ref{12}), we have
$$
K=\frac{12}{\pi^2}\int\limits_{0}^{q_0^2} dq^2\,q^2\left(
\frac{C}{q^2}+z-h_0 \right)=\frac{12}{\pi^2}q_0^4\left(
\frac{C}{q_0^2}+\frac{1}{2}(z-h_0) \right) =
$$
\begin{equation}
=\frac{6}{\pi^2}(h_0-z)x_0^2\Lambda^4_{QCD}.
\label{13}
\end{equation}
Expression~(\ref{13}) is valid for each of 1 --- 4-loop
approximations of the perturbative $\alpha_s$ and 
ratio  $K/\Lambda^4_{QCD}$ does not depend on a 
normalization of $\alpha_s$. In the one-loop case we have 
from~(\ref{6}),
(\ref{8}), (\ref{13}) with $b_0=$ 9 
\begin{equation}
K=\frac{e^4}{6\pi^2}\Lambda^4_{QCD}.
\label{14}
\end{equation}
In the one-loop case normalization condition 
$\alpha_s(m^2_{\tau})=$ 0.32 gives $\Lambda_{QCD}=$ 0.201 GeV, and
expression~(\ref{14}) gives $K=$ (0.197 GeV)$^4$, which is far from
the conventional value of the gluon 
condensate (0.33 $\pm$ 0.01 GeV)$^4$~\cite{Shif}. 

Let us consider multi-loop cases. Solution $h(q^2)$ of 
equation~(\ref{2}) for $L=\ln(q^2/\Lambda^2)\rightarrow
\infty$ reads as follows  
$$
h(q^2)=\frac{1}{b_0 L}\left\{1-\frac{2b_1}{b_0
^2L}\ln L+\frac{4b_1^2}{b_0^4L^2}
\left[\ln^2 L-\ln L-1+\frac{b_0
b_2}{8b_1^2}\right]\right.
$$
$$
-\frac{8b_1^3}{b_0^6L^3}\left[
\ln^3 L
-\frac{5}{2}\ln^2 L-\left(2-\frac{3b_0b_2}
{8b_1^2}\right)\ln L\right.
$$
\begin{equation}
\left.\left.+\frac{1}{2}-\frac{b_0^2
b_3}{16b_1^3}\right]+O\left(\frac{1}{L^4}\right)\right\}.
\label{15}
\end{equation}
Keeping in the expression terms with powers of logarithms 
in denominators up to the first, the second, the third 
and the fourth, we fix the 1 --- 4-loop approximations of 
the perturbation theory for running coupling constant. 
It may be written in the form
$$
\alpha_s(q^2)
=4\pi h(q^2)
=\frac{4\pi}{b_0}a(x),
$$
$$
a(x)=\frac{1}{\ln x}-
b\frac{\ln(\ln x)}{\ln^2x}+b^2\left[\frac{\ln^2(\ln x)}{\ln^3x}-
\frac{\ln(\ln x)}{\ln^3x}+\frac{\kappa}{\ln^3x}\right]
$$
\begin{equation}
-b^3\left[\frac{\ln^3(\ln
x)}{\ln^4x}-\frac{5}{2}\frac{\ln^2(\ln x)}{\ln^4x}+(3\kappa+1)
\frac{\ln(\ln x)}{\ln^4x}+\frac{\bar\kappa}{\ln^4 x}\right].
\label{16}
\end{equation}
Here $x=q^2/\Lambda^2$, and coefficient are defined as 
follows
\begin{eqnarray}
b&=&\frac{2b_1}{b^2_0},\nonumber\\
\kappa&=&-1+\frac{b_0 b_2}
{8b^2_1},\nonumber\\
\bar\kappa&=&
\frac{1}{2}-\frac{b_0^2b_3}{16b_1^3}.
\label{17}
\end{eqnarray}
Coefficients  $b$, $\kappa$, $\bar\kappa$ depend on 
$n_f$.With $n_f=$ 3 we have $b\simeq$ 0.7901, $\kappa 
\simeq$ 0.4147, $\bar\kappa \simeq -1.3679$.
In the case of the two-loop approximation for 
perturbative  $\alpha_s$ we have the following set of 
equations 
$$
b_0 h_0^2+2b_1h_0^3=h_0-z,
$$
$$
2b_0 h_0+6b_1h_0^2=1,
$$
\begin{equation}
\frac{1}{\ln x_0}-
b\frac{\ln(\ln x_0)}{\ln^2x_0}
=b_0h_0,
\label{18}
\end{equation}
$$
\frac{c_0}{x_0}+z=h_0.
$$
The set for three loops reads
$$
b_0 h_0^2+2b_1h_0^3+\frac{b_2}{2}h_0^4=h_0-z,
$$
$$
2b_0 h_0+6b_1h_0^2+2b_2h_0^3=1,
$$
\begin{equation}
\frac{1}{\ln x_0}-
b\frac{\ln(\ln x_0)}{\ln^2x_0}+b^2\left[\frac{\ln^2(\ln
x_0)}{\ln^3x_0}-
\frac{\ln(\ln x_0)}{\ln^3x_0}+\frac{\kappa}{\ln^3x_0}\right]
=b_0h_0,
\label{19}
\end{equation}
$$
\frac{c_0}{x_0}+z=h_0.
$$
The set for four loops reads
$$
b_0 h_0^2+2b_1h_0^3+\frac{b_2}{2}h_0^4+b_3h_0^5=h_0-z,
$$
$$
2b_0 h_0+6b_1h_0^2+2b_2h_0^3+5b_3h_0^4=1,
$$
$$
\frac{1}{\ln x_0}-
b\frac{\ln(\ln x_0)}{\ln^2x_0}+b^2\left[\frac{\ln^2(\ln
x_0)}{\ln^3x_0}-
\frac{\ln(\ln x_0)}{\ln^3x_0}+\frac{\kappa}{\ln^3x_0}\right]
$$
\begin{equation}
-b^3\left[\frac{\ln^3(\ln
x_0)}{\ln^4x_0}-\frac{5}{2}\frac{\ln^2(\ln
x_0)}{\ln^4x_0}+(3\kappa+1)
\frac{\ln(\ln x_0)}{\ln^4x_0}+\frac{\bar\kappa}{\ln^4
x_0}\right]=b_0h_0,
\label{20}
\end{equation}
$$
\frac{c_0}{x_0}+z=h_0.
$$
From sets of equations~(\ref{18}) -- (\ref{20}) we find 
values of $h_0$, $z$, $x_0$ and $á_0$. They are presented 
in Table~\ref{tab1}. The value of $K^{1/4}/\Lambda_{QCD}$ 
calculated with the aid of expression~(\ref{13}) is also 
presented there. For the parameter of string tension 
$\sigma$ by using expressions~(\ref{9}), (\ref{10}) we have 
relation $\sigma/\Lambda_{QCD}^2 = 8 \pi c_0/3$. 
Taking into account the existing 
data~\cite{Data,Bethke,Piv} we fix the momentum dependence 
of solutions 
\begin{table}[!ht]
\caption{
The dimensionless parameters 
$h_0$, $z$, $x_0=q_0^2/\Lambda_{QCD}^2$,
$c_0=C/\Lambda_{QCD}^2$, $K^{1/4}/\Lambda_{QCD}$ 
on the number of loops,  
$n_f=$ 3}
\label{tab1}
\begin{center}
\begin{tabular}{|l| r| r| r| r|}\hline 
  &1-loop &2-loop&3-loop&4-loop
	 \\ \hline
$h_0$  
 &   0.0556 &   0.0392 &   0.0356 &   0.0337
\\ \hline
$z$ 
 &   0.0278 &   0.0215 &   0.0203 &   0.0197
\\ \hline 
$x_0$  
 &   7.3891 &   7.7763 &  10.2622 &  12.4305
\\ \hline 
$c_0$  
 &   0.2053 &   0.1374 &   0.1572 &   0.1741
\\ \hline 
$K^{1/4}/\Lambda_{QCD}$
 &   0.9799 &   0.8977 &   0.9952 &   1.0709
\\ \hline
\end{tabular}
\end{center}
\end{table}
for a number of values of the running coupling constant 
at $\tau$-lepton mass scale $m_\tau$ with the effective 
number of flavors $n_f$ being equal to 3. 
The corresponding to these normalization conditions values 
of $\Lambda_{QCD}$ are presented in Table~\ref{tab2}, values of 
boundary momentum $q_0=$ $\sqrt{x_0}\Lambda_{QCD}$ are 
presented in Table~\ref{tab3}, values of the gluon condensate 
$K^{1/4}$ are presented in Table~\ref{tab4} and the string 
tension parameter $\sigma$ is given in Table~\ref{tab5}. 
\begin{table}[!ht]
\caption{
Values of the parameter $\Lambda_{QCD}$ (GeV)
on loop numbers and normalization 
conditions. Normalization conditions:
$\alpha_s(m_{\tau}^2)=$ 0.29, 0.30, ..., 0.36 with 
$m_{\tau}=$ 1.77703 GeV, $n_f=$ 3}
\label{tab2}
\begin{center}
\begin{tabular}{|l| r| r| r| r|}\hline 
$\alpha_s(m_{\tau}^2)$  &1-loop &2-loop&3-loop&4-loop
	 \\ \hline
0.29  
 &   0.1600 &   0.3168 &   0.2873 &   0.2837
\\ \hline
0.30 
 &   0.1734 &   0.3370 &   0.3069 &   0.3026
\\ \hline 
0.31  
 &   0.1869 &   0.3568 &   0.3263 &   0.3212
\\ \hline 
0.32  
 &   0.2005 &   0.3762 &   0.3454 &   0.3394
\\ \hline
0.33
 &   0.2143 &   0.3951 &   0.3642 &   0.3573
\\ \hline
0.34
 &   0.2280 &   0.4136 &   0.3827 &   0.3749
\\ \hline
0.35
 &   0.2418 &   0.4315 &   0.4007 &   0.3920
\\ \hline
0.36
 &   0.2556 &   0.4490 &   0.4184 &   0.4087
\\ \hline 
\end{tabular}
\end{center}
\end{table}
\begin{table}[!ht]
\caption{
Values of the parameter $q_0$ (GeV)  on loop 
numbers and normalization 
conditions. Normalization conditions are the same as in
Table~\ref{tab2}}
\label{tab3}
\begin{center}
\begin{tabular}{|l| r| r| r| r|}\hline 
$\alpha_s(m_{\tau}^2)$  &1-loop &2-loop&3-loop&4-loop
	 \\ \hline
0.29  
 &   0.4350 &   0.8833 &   0.9203 &   1.0002
\\ \hline
0.30 
 &   0.4713 &   0.9397 &   0.9831 &   1.0667
\\ \hline 
0.31  
 &   0.5081 &   0.9949 &   1.0452 &   1.1323
\\ \hline 
0.32  
 &   0.5451 &   1.0490 &   1.1065 &   1.1967
\\ \hline
0.33
 &   0.5824 &   1.1018 &   1.1667 &   1.2598
\\ \hline
0.34
 &   0.6198 &   1.1533 &   1.2258 &   1.3216
\\ \hline
0.35
 &   0.6572 &   1.2034 &   1.2838 &   1.3820
\\ \hline
0.36
 &   0.6947 &   1.2522 &   1.3405 &   1.4409
\\ \hline 
\end{tabular}
\end{center}
\end{table}
\begin{table}[!ht]
\caption{
Values of the gluon condensate $K^{1/4}$ (GeV)  on loop 
numbers and normalization 
conditions. Normalization conditions are the same as in
Table~\ref{tab2}}
\label{tab4}
\begin{center}
\begin{tabular}{|l| r| r| r| r|}\hline 
$\alpha_s(m_{\tau}^2)$  &1-loop &2-loop&3-loop&4-loop
	 \\ \hline
0.29  
 &   0.1568 &   0.2844 &   0.2859 &   0.3038
\\ \hline
0.30 
 &   0.1699 &   0.3025 &   0.3054 &   0.3240
\\ \hline 
0.31  
 &   0.1832 &   0.3203 &   0.3247 &   0.3439
 \\ \hline 
0.32  
 &   0.1965 &   0.3377 &   0.3437 &   0.3635
\\ \hline
0.33
 &   0.2099 &   0.3547 &   0.3624 &   0.3827
\\ \hline
0.34
 &   0.2234 &   0.3713 &   0.3808 &   0.4014
\\ \hline
0.35
 &   0.2369 &   0.3874 &   0.3988 &   0.4198
\\ \hline
0.36
 &   0.2504 &   0.4031 &   0.4164 &   0.4377
\\ \hline 
\end{tabular}
\end{center}
\end{table}
\begin{table}[!ht]
\caption{
String tension parameter $a=\sqrt{\sigma}$ (GeV)  on loop
numbers and normalization 
conditions. Normalization conditions are the same as in
Table~\ref{tab2}}
\label{tab5}
\begin{center}
\begin{tabular}{|l| r| r| r| r|}\hline 
$\alpha_s(m_{\tau}^2)$  &1-loop &2-loop&3-loop&4-loop
	 \\ \hline
0.29  
 &   0.2098 &   0.3398 &   0.3297 &   0.3426
\\ \hline
0.30 
 &   0.2274 &   0.3615 &   0.3522 &   0.3654
\\ \hline 
0.31  
 &   0.2451 &   0.3827 &   0.3745 &   0.3878
 \\ \hline 
0.32  
 &   0.2630 &   0.4035 &   0.3964 &   0.4099
\\ \hline
0.33
 &   0.2809 &   0.4239 &   0.4180 &   0.4315
\\ \hline
0.34
 &   0.2990 &   0.4437 &   0.4392 &   0.4527
\\ \hline
0.35
 &   0.3170 &   0.4629 &   0.4599 &   0.4733
\\ \hline
0.36
 &   0.3351 &   0.4817 &   0.4802 &   0.4935
\\ \hline 
\end{tabular}
\end{center}
\end{table}

Let us consider another variant of definition of 
the perturbative $\alpha_s$ behavior for $q^2<q^2_0$.  Namely
instead of freezing~(\ref{12}) we shall assume 
``forced'' analytic behavior of $\alpha_s$ in this region,  
\begin{equation}       
\alpha_{\rm pert}(q^2)\,=\, 
\alpha_{\rm an}(q^2), \;\;\;  q^2<q_0^2.
\label{21}
\end{equation}
The main ideas of the analytic approach in  quantum field 
theory, which allows one to overcome difficulties, connected 
with nonphysical singularities in perturbative 
expressions, were proposed in works~\cite{Red,Bog}. The 
analytic approach is successfully applied to QCD
~\cite{SolShirTMF}. The forced analytic running coupling 
constant is defined by the spectral representation
\begin{equation}       
a_{\rm an}(y)=\frac{1}{\pi}\int\limits_0^\infty
\frac{d\sigma}{y+\sigma}
\rho(\sigma),
\label{22}
\end{equation}
where spectral density $\rho(\sigma)=
\Im a_{\rm an}(-\sigma-i0)=$
$\Im a(-\sigma-i0)$. For the perturbative solutions $a(x)$ 
of the form~(\ref{16}) the two-loop analytic 
running coupling constant and its non-perturbative part 
were studied in ref.~\cite{PR}, the three-loop case 
and the four-loop case where studied in
refs.~\cite{YadFiz} and~\cite{A4}, 
respectively. 
Let us write the spectral density up to the four-loop 
case.
\begin{equation}
\rho^{(1)}(\sigma)=
\frac{\pi}{t^2+\pi^2},
\label{23}
\end{equation}
\begin{equation}
\rho^{(2)}(\sigma)=\rho^{(1)}(\sigma)
-\frac{b}{(t^2+\pi^2)^2}\left[
2\pi t F_1(t)-\left(t^2-\pi^2\right)F_2(t)\right],
\label{24}
\end{equation}
$$
\rho^{(3)}(\sigma)=\rho^{(2)}(\sigma)+\frac{b^2}{(t^2+\pi^2)^3}
\left[\pi
\left(3t^2-\pi^2\right)\left(
F_1^2(t)-F_2^2(t)\right)-
2t\left(t^2-3\pi^2\right)F_1(t)F_2(t)\right.
$$
\begin{equation}
-\pi\left(3t^2-\pi^2\right)F_1(t)
\left.+t\left(t^2-3\pi^2\right)F_2(t)
+\pi\kappa\left(3t^2-\pi^2\right)\right],
\label{25}
\end{equation}
$$
\rho^{(4)}(\sigma)=\rho^{(3)}(\sigma)-\frac{b^3}
{(t^2+\pi^2)^4}\left[\left(t^4-
6\pi^2t^2+\pi^4\right)\left(F_2^3(t)-3F_1^2(t)F_2(t)
\right)
\right.
$$
$$
+4\pi t\left(t^2-
\pi^2\right)\left(F_1^3(t)-3F_1(t)F_2^2(t)\right)
-10\pi t\left(t^2-\pi^2\right)\left(F_1^2(t)-F_2^2(t)
\right)
$$
$$
+5\left(t^4-
6\pi^2t^2+\pi^4\right)F_1(t)F_2(t)
+4\pi \left(1+3\kappa\right)t\left(
t^2-\pi^2\right)F_1(t)
$$
\begin{equation}
-\left.\left(1+3\kappa\right)
\left(t^4-6\pi^2t^2+\pi^4\right)F_2(t)+
4\pi \bar\kappa t\left(t^2-\pi^2\right)\right].
\label{26}
\end{equation}
Here $t=\ln(\sigma)$, 
\begin{equation}
F_1(t)\equiv\frac{1}{2}\ln(t^2+\pi^2),\,\,\,
F_2(t)\equiv\arccos\frac{t}{\sqrt{t^2+\pi^2}}.
\label{27}
\end{equation}
Solving the equation\footnote{While solving this equation 
it is convenient to use the method of ref.~\cite{YadFiz}, 
in which $\alpha_{\rm an}^{\rm npt}(y)$ is represented as a series 
in inverse powers of $y$} 
\begin{equation}
a_{\rm an}(y)=b_0 h_0,
\label{28}
\end{equation}
we find values $y_0$ for $a_{\rm an}$, being defined by 
formulas~(\ref{22}) -- (\ref{27}) with the use of values 
$h_0$ obtained above.
Further we find dimensionless quantity  $\xi=$
$(\Lambda_{\rm an}/\Lambda_{QCD})^2=$ $x_0/y_0$. Values of 
$y_0$ and $\xi$ in dependence on the number of loops are 
presented in Table~\ref{tab6}.
In Table~\ref{tab7} values of $\Lambda_{\rm an}=$
$q_0/\sqrt{y_0}$ are presented in dependence on the 
number  of loops and on normalization conditions at 
$q^2=m_{\tau}^2$. 

Let us turn to the gluon condensate. In the  
considered method of definition of the perturbative 
part of $\alpha_s$ in the non-perturbative region we have
\begin{equation}
K_{\rm an}=
\frac{3}{\pi^3}\int\limits_{0}^{q_0^2} dq^2\,q^2(\alpha_s(q^2)-
\alpha_{\rm an}(q^2))=\frac{12}{\pi^2}\int\limits_{0}^{q_0^2}
dq^2\,q^2\left(
\frac{C}{q^2}+z-\frac{1}{b_0}a_{\rm an}\left(\frac{q^2}
{\Lambda_{\rm an}^2}
\right) \right).
\label{29}
\end{equation}
Taking into account the  spectral representation~(\ref{22}) 
and performing integration in Eq.~(\ref{29}), we obtain
\begin{equation}
K_{\rm an}=
\frac{12}{\pi^2}\left(C q_0^2+\frac{z}{2}q_0^4\right)
-\frac{12\Lambda^4_{\rm an}}{\pi^3 b_0}\int\limits_{0}^{\infty}
d\sigma\,\rho(\sigma)\left[y_0-\sigma\ln\left(1+\frac{y_0}
{\sigma}
\right)\right].
\label{30}
\end{equation}
Substitution $\sigma=\exp(t)$ with the use of expressions
~(\ref{8}) and (\ref{10}) leads to the following 
representation of the gluon condensate
\begin{equation}
K_{\rm an}=
\Lambda_{QCD}^4\left[\frac{12}{\pi^2}\left(h_0-\frac{z}{2}\right)
x_0^2
-\frac{12\xi^2}{\pi^3 b_0}\int\limits_{-\infty}^{\infty}
d t\,\rho(t)\left\{y_0e^t-e^{2t}\ln\left(1+y_0e^{-t}\right)\right\}
\right].
\label{31}
\end{equation}
The expression in the square brackets of~(\ref{31}) does 
not depend on values of $\alpha_s(m_{\tau}^2)$, values of 
the ratio $K_{\rm an}^{1/4}/\Lambda_{QCD}$, which are 
obtained by numerical integration in formula~(\ref{31}), 
are given in Table~\ref{tab6}.
In Table~\ref{tab7} and Table~\ref{tab8} values of the parameter
$\Lambda_{\rm an}$ and of the gluon condensate  
$K_{\rm an}^{1/4}$ are presented, respectively. 
\begin{table}[!ht]
\caption{
Dimensionless parameters  $y_0$, $\xi$, 
$K_{\rm an}^{1/4}/\Lambda_{QCD}$ 
on  
loop numbers, $n_f=$ 3}
\label{tab6}
\begin{center}
\begin{tabular}{|l| r| r| r| r|}\hline 
  &1-loop &2-loop&3-loop&4-loop
	 \\ \hline
$y_0$  
 &   1.0000 &   0.8299 &   2.2691 &   2.8710
\\ \hline
$\xi$ 
 &   7.3893 &   9.3702 &   4.5225 &   4.3297
\\ \hline 
$K_{\rm an}^{1/4}/\Lambda_{QCD}$  
 &   0.9373 &   0.8494 &   0.9353 &   1.0025
\\ \hline 
\end{tabular}
\end{center}
\end{table}
\begin{table}[!ht]
\caption{
Dependence of the parameter $\Lambda_{\rm an}$ (GeV) 
on loop 
numbers and normalization 
conditions. Normalization conditions are the same as in
 Table~\ref{tab2}}
\label{tab7}
\begin{center}
\begin{tabular}{|l| r| r| r| r|}\hline 
$\alpha_s(m_{\tau}^2)$  &1-loop &2-loop&3-loop&4-loop
	 \\ \hline
0.29  
 &   0.4350 &   0.9696 &   0.6109 &   0.5903
\\ \hline
0.30 
 &   0.4714 &   1.0315 &   0.6526 &   0.6296
\\ \hline 
0.31  
 &   0.5081 &   1.0921 &   0.6939 &   0.6682
\\ \hline 
0.32  
 &   0.5451 &   1.1515 &   0.7345 &   0.7063
\\ \hline
0.33
 &   0.5824 &   1.2094 &   0.7745 &   0.7435
\\ \hline
0.34
 &   0.6198 &   1.2659 &   0.8138 &   0.7800
\\ \hline
0.35
 &   0.6572 &   1.3210 &   0.8522 &   0.8156
\\ \hline
0.36
 &   0.6947 &   1.3745 &   0.8899 &   0.8504
\\ \hline 
\end{tabular}
\end{center}
\end{table}
\begin{table}[!ht]
\caption{
Values of the gluon condensate $K_{\rm an}^{1/4}$ (GeV)
in dependence on loop 
numbers and normalization 
conditions. Normalization conditions are the same as in
 Table~\ref{tab2}}
\label{tab8}
\begin{center}
\begin{tabular}{|l| r| r| r| r|}\hline 
$\alpha_s(m_{\tau}^2)$  &1-loop &2-loop&3-loop&4-loop
	 \\ \hline
0.29  
 &   0.1500 &   0.2691 &   0.2687 &   0.2844
\\ \hline
0.30 
 &   0.1625 &   0.2862 &   0.2870 &   0.3033
\\ \hline 
0.31  
 &   0.1752 &   0.3031 &   0.3052 &   0.3219
 \\ \hline 
0.32  
 &   0.1880 &   0.3195 &   0.3230 &   0.3403
\\ \hline
0.33
 &   0.2008 &   0.3356 &   0.3406 &   0.3582
\\ \hline
0.34
 &   0.2137 &   0.3513 &   0.3579 &   0.3758
\\ \hline
0.35
 &   0.2266 &   0.3666 &   0.3748 &   0.3929
\\ \hline
0.36
 &   0.2395 &   0.3814 &   0.3914 &   0.4097
\\ \hline 
\end{tabular}
\end{center}
\end{table}

\begin{figure}[!ht]
\centerline{\psfig{figure=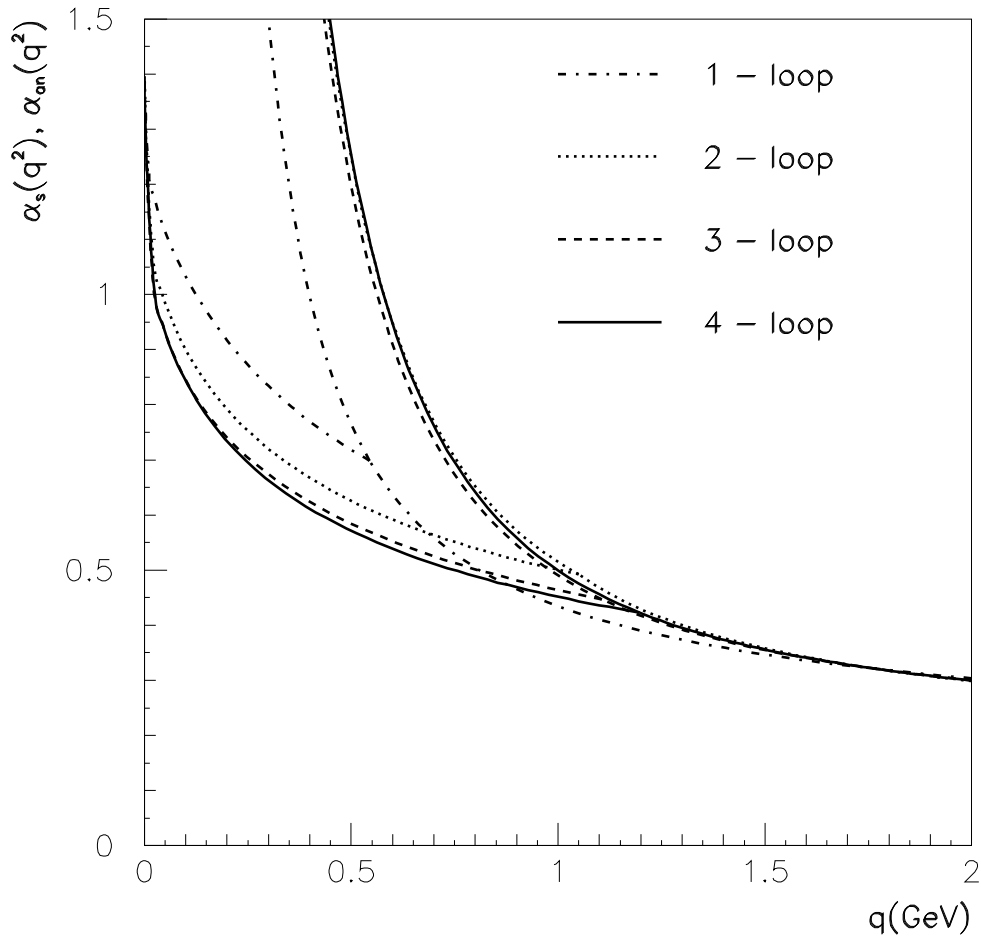,height=15cm,width=15cm}}
\caption{Running coupling constant $\alpha_s(q^2)$ for 
all $q^2>0$ and analytic coupling constant 
$\alpha_{\rm an}(q^2)$ for $q^2<q^2_0$ (the corresponding curves 
are the lower ones). Normalization conditions:
$\alpha_s(m_{\tau}^2)=$ 0.32, $\alpha_{\rm an}(q_0^2)=$ $\alpha_s
(q_0^2)$
}
\label{fig1}
\end{figure}

\begin{figure}[!ht]
\centerline{\psfig{figure=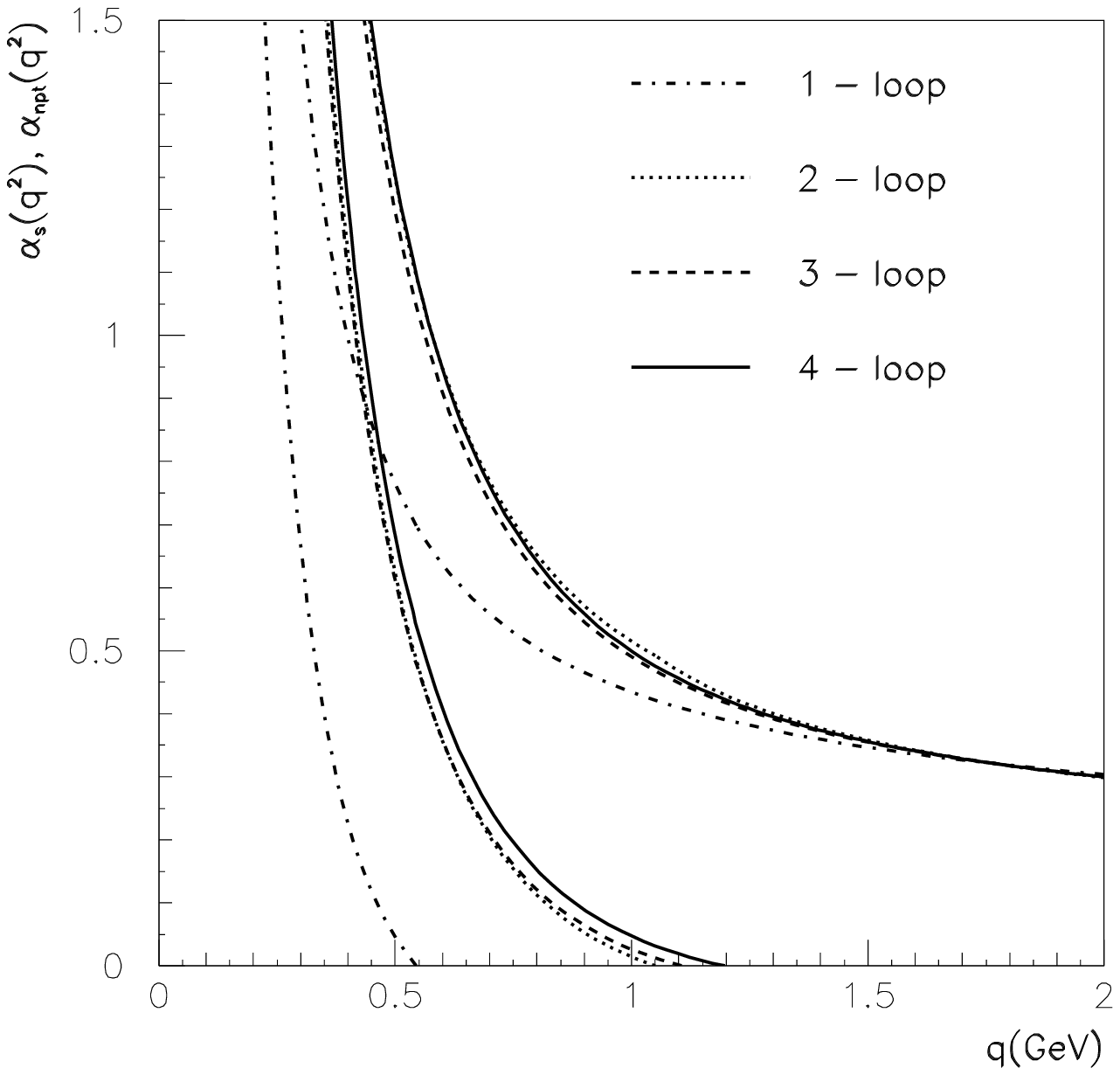,height=15cm,width=15cm}}
\caption{Running coupling constant $\alpha_s(q^2)$ and 
its non-perturbative part 
$\alpha_{\rm npt}(q^2)$ (the corresponding curves 
are the lower ones) with definition of 
$\alpha_{\rm pert}(q^2)$ by (\ref{21}). 
Normalization condition: $\alpha_s(m_{\tau}^2) = 0.32$
}
\label{fig2}
\end{figure}

For cases from one loop up to four loops the behavior of the
running coupling constant $\alpha_s(q^2)$ for all $q^2>0$ is 
shown in~Fig.~\ref{fig1}. Here the behavior of the analytic 
coupling constant $\alpha_{\rm an}(q^2)$ for $q^2<q^2_0$, which 
defines the perturbative part of $\alpha_s(q^2)$ in this 
region, is also shown. In~Fig.~\ref{fig2} in addition to 
$\alpha_s(q^2)$ its non-perturbative part 
$\alpha_{\rm npt}(q^2)$ is also shown, which turns to zero for 
$q^2 > q_0^2$. Normalization condition for all curves 
in~ Fig.~\ref{fig1},  Fig.~\ref{fig2} is $\alpha_s(m_\tau^2) = 
0.32$.
%
%
%
%

Starting from the well-known perturbative behavior of 
the $\beta$-function for small values of the coupling 
constant $h$ and the behavior for large values of the coupling 
constant, which corresponds to the linear confinement, we 
have constructed the $\beta$-function for all $h > 0$. 
To control dependence of the results on the number of loops we 
simultaneously consider cases corresponding to perturbative 
$\beta$-function for 1 --- 4 loops. While constructing  
the matched $\beta$-function we assume it to be smooth at 
the matching point, that leads to the invariant charge 
being smooth together with its derivative for all 
$q^2 > 0$. The normalization of the invariant charge, e.g., 
at  $m_\tau$ fix it thoroughly. Then value of  $\alpha_s(m_\tau^2)
\simeq 0.33$ corresponds to
string tension $a \simeq 0.42$ GeV~\cite{Sol}.

The obtained invariant charge is applied to  study of 
a quite important physical quantity, the gluon condensate. 
In doing this we consider two variants of extracting of the
non-perturbative contributions from the overall expression 
for the invariant charge. The first variant assumes 
``freezing" of perturbative part of the charge in the 
non-perturbative region $q^2 < q_0^2$, while for the 
second one we choose the analytic behavior of the 
perturbative part in this region. As we see from Tables~
\ref{tab4}, \ref{tab8} the first variant leads to 
values of the gluon condensate being somewhat larger, than 
that for the second variant. Emphasize, that for 
$\alpha_s(m_\tau^2) = 0.33$ the gluon condensate for 
the second variant practically coincides with the 
conventional value~\cite{Shif}, while for the first 
variant it turns to be $K^{1/4} \simeq 0.36 $ GeV, that is 
rather higher, than the conventional value. Note, that 
other important non-perturbative parameter $q_0$ for the 
same normalization conditions also turns to be of  
reasonable magnitude, $q_0 \simeq 1.17$ GeV 
(Table~\ref{tab3}). We may conclude, that singular 
behavior of the running coupling constant~(\ref{3}) 
consistently describe these non-perturbative 
parameters.

The authors express deep gratitude to V.A. Petrov, 
V.E. Rochev and D.V. Shirkov for valuable discussions. 
The work has been partially supported by RFBR under 
Grant No. 02-01-00601.


\begin{thebibliography}{99}
\bibitem{ShirTMF02}
D.V.~Shirkov, Teor. Mat. Fiz. {\bf 132},
484 (2002) 
[Theor. Math. Phys. {\bf 132}, 1309 (2002)].  
\bibitem{Arb1}
B.A.~Arbuzov, Fiz. Elem. Chast. Atom. Yad. {\bf 19}, 5 
(1988) [Phys. Part. Nucl. {\bf 19}, 1 (1988)].
\bibitem{AA} A.I. Alekseev and B.A. Arbuzov, Yad. Fiz. 
{\bf 61}, 314 (1998) [Phys. At. Nucl. {\bf 61}, 264 
(1998)]; A.I. Alekseev and B.A. Arbuzov, Mod. Phys. 
Lett. A, {\bf 13}, 1447 (1998); A.I. Alekseev, in 
{\it Proceedings of the Workshop on Non-perturbative 
Methods in Quantum Field Theory}, Adelaide, Australia, 
1998, edited by A.V. Schreiber, A.G. Williams and A.W. 
Thomas (World Scientific, Singapore, 1998), 
hep-ph/9808206.
\bibitem{lat}
G. Burgio, F.Di Renzo, G. Marchesini and E. Onofri, 
Phys. Lett. B {\bf 422}, 219 (1998), Nucl. 
Phys. Proc. Suppl. {\bf 63}, 805 (1998);
G. Burgio, F.Di Renzo, C. Parrinello and C. Pittori, 
 Nucl. Phys. Proc. Suppl. {\bf 73}, 623 (1999), Nucl. 
Phys. Proc. Suppl. {\bf 74}, 388 (1999); Ph. Boucaud et 
al., JHEP {\bf 0004}, 006 (2000).
\bibitem{Gogohia}
V. Gogohia, Phys. Lett. B {\bf 584}, 225 (2004).
\bibitem{Data}
Particle Data Group, D.E.~Groom {\it et al.}, Eur. Phys. 
J. C {\bf 15}, 85
(2000).
\bibitem{Simonov}
Yu.A.~Simonov,
Yad. Fiz. {\bf 58}, 113  (1995) [Phys. At. Nucl.
{\bf 58},107 (1995]; Yad. Fiz. {\bf 65}, 140  (2002).
\bibitem{Shif} M.A. Shifman, A.I. Vainshtein and V.I. 
Zakharov, Nucl. Phys. B {\bf 147}, 385, 448 (1979).
\bibitem{Bethke}
S.~Bethke, J. Phys. G {\bf 26}, R27 (2000),
 hep-ex/0004021.
\bibitem{Piv}
A.A. Pivovarov, Preprint INR-TH-03-3, Moscow 2003, 
hep-ph/0301074.
\bibitem{Red}
P.J.~Redmond,  Phys. Rev.  {\bf 112}, 1404 (1958).
\bibitem{Bog}
N.N.~Bogolubov, A.A.~Logunov, and  D.V.~Shirkov, 
Zh. \'{E}ksp. Teor.Fiz. {\bf 37}, 805 (1959) 
[Sov. Phys. JETP
{\bf 10}, 574 (1960)].
\bibitem{SolShirTMF}
I.L.~Solovtsov and D.V.~Shirkov, Teor. Mat. Fiz. {\bf 120},
482 (1999) 
[Theor. Math. Phys. {\bf 120}, 1210 (1999)].
\bibitem{PR} A.I.~Alekseev, Phys. Rev. D, {\bf 61}, 114005 
(2000).
\bibitem{YadFiz}
A.I.~Alekseev,  
Yad. Fiz. {\bf 65}, 1722  (2002) [Phys. At. Nucl.
{\bf 65},1678 (2002)].
\bibitem{A4} A.I.~Alekseev, Few-Body Systems {\bf 32}, 193 
(2003).
\bibitem{Sol} L.D. Soloviev, Phys. Rev. D {\bf 58}, 035005 
(1998); Phys. Rev. D {\bf 61}, 015009 (2000). 
\end{thebibliography}
\end{document}